# Bit Error Rate Analysis of M-ary PSK and M-ary QAM over Rician Fading Channel

[1]Subrato Bharati, [2]Mohammad Atikur Rahman, [3]Prajoy Podder, [4]Muhammad Ashiqul Islam,[5]Mohammad Hossain

[1,2,3,5] Department of EEE, Ranada Prasad Shaha University, Narayanganj, Bangladesh.

[3,4] Department of ECE, Khulna University of Enginnering & Technology, Khulna, Bangladesh.

[1]subratobharati1@gmail.com, [2]sajibextreme@gmail.com, [3]prajoypodder@gmail.com, [4]kuet0909018@hotmail.com, [5]mohammadandhossain@gmail.com,

*Abstract*—This paper mainly illustrates the Bit error rate performance of M-ary QAM and M-ary PSK for different values of SNR over Rician Fading channel. A signal experiences multipath propagation in the wireless communication system which causes expeditious signal amplitude fluctuations in time, is defined as fading. Rician Fading is a small signal fading. Rician fading is a hypothetical model for radio propagation inconsistency produced by fractional cancellation of a radio signal by itself and as a result the signal reaches in the receiver by several different paths. In this case, at least one of the destination paths is being lengthened or shortened. From this paper , it can be observed that the value of Bit error rate decreases when signal to noise ratio increases in decibel for M-ary QAM and M-ary PSK such as 256 QAM, 64 PSK etc. Constellation diagrams of M-QAM and M-PSK have also been showed in this paper using MATLAB Simulation. The falling of Bit error rate with the increase of diversity order for a fixed value of SNR has also been included in this paper. Diversity is a influential receiver system which offers improvement over received signal strength.

*Keywords— M-ary PSK, QAM, Bit error rate, SNR, constellation diagrams, Rician fading channel.*

## I. INTRODUCTION

Last few decades people communicate with the remote people with the help of wire connection. From few years ago, people increase their attention about wireless movable communication system and then the day has come. Present day is the time of wireless communication. The biggest part of communication is growing by wireless technologies. The performance of the good transmitting and receiving systems is very important. For that, many digital modulation and demodulation process are available nowadays. Like other modulation techniques QAM and M-ary PSK is also the technique of digital modulation. In PSK each signal represents 1 bit but in M-ary PSK each phase represents n bits. Where n depend on the number of phases. The advantages of M-ary PSK are, as with BPSK, there are phase ambiguity problems at the receiver and differentially encoded M-ary PSK is more normally used in practice. It is use of differential M-ary PSK where the phase shift is not relation to a reference signal but only compares multiple signal to rebuild data. Disadvantages are, inter channel interference is expressively large in M-ary PSK. The disadvantage of M-ary PSK relative to BPSK is that it is more sensitive to phase variations. Efficient usage of bandwidth is the major benefit of QAM modulation deviations. This is due to the fact that QAM symbolize more number of bits per carrier. Disadvantages are, QAM modulation process is more prehensile to the noise. Due to this QAM, receiver is more difficult compare to receivers of other modulation types. To transmit QAM signal linear amplifier is needed which consumes more power. After that, M-ary PSK is harder to implement if the phase is relatively small. It is preferred to use an IQ modem. Finally, they are both the same theoretically but the implementation is different. So we prefer QAM.

In wireless communication, fading of the signals received by the mobile unit is an inherent problem. Since the mobile unit rapidly changes its location and keeps on changing with the mobile phone users, the resultant radio signal incident on its antenna varies continuously. Slow and fast are the two aspects of multipath propagation. In slow fading, propagation changes are slow due to the small amount of relatiove motion between mobile and diffracting objects. On the other hand fast fading is the rapid fluctuations of the received signal due to the sudden movement of mobile [3].

Jinhua Lu [4] discussed signal space concepts to evaluate the BER performance of M-ary PSK and M-QAM over AWGN Channel. S. Chennakeshu [5] provided a unified method to derive the exact symbol error rate and bit error rate for MPSK signals considering N channel diversity reception for Rayleigh Fading. M.O. Hasna [6] discuss about average bit-error rate expressions for differential binary phase-shift keying, and as well as probability of outage formulas for noise limited systems are run. He provided a method of transmission systems which relays over Rayleigh-fading channels in end-to-end performance of transmission systems with relays over Rayleigh-fading channels. The main difference between Rayleigh fading and Rician channel is that Line of sight signal does not exist in Rayleigh fading but exists in Rician fading channel. When there is dominant non fading signal component present, the small signal fading enveloped can be denoted as

Rician fading. BER performance of Rician fading channel has been showed here with simulation.

## II. THEORETICAL DESCRIPTION

### A. M-ary PSK

Phase shift keying is a digital modulation process which carries data by changing the phase of the carrier wave. There are several methods that can be used accomplish PSK. In binary phase shift keying to opposite signal phases are used because there are two possible wave phase. The digital signal is separated time wise into individual bits. For the different bits phase will be changed for the two same values phase will be unchanged. Binary phase shift keying is sometimes called be phase modulation. M-ary phase shift keying technic is simpler to binary phase shift keying. M-ary PSK or MPSK is the type of phase shift keying where variations of multiple phase are used instead of two phases [7]. In PSK is signal represents one bit but in M-ary PSK is phase represents n bits where n depends on the number of phases.

M-ary PSK modulated signal,

$$S(t) = A_c \cos(\omega_c t + n\pi/4) \quad ; 0 \leq t \leq T \quad \ldots \ldots (1)$$

Where, n = 1, 2, 3, 4, …..etc.

Fig. 1 and 2 show the constellation diagram of 16 PSK and 64 PSK whose Phase offset is 0.19635 rad and 0.049087 rad respectively.

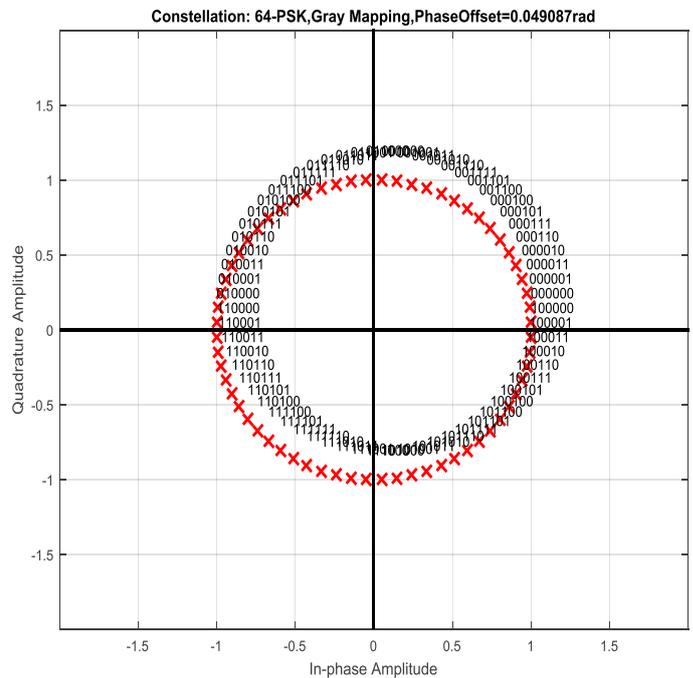

Fig.2. Constellation Diagram of 64-PSK

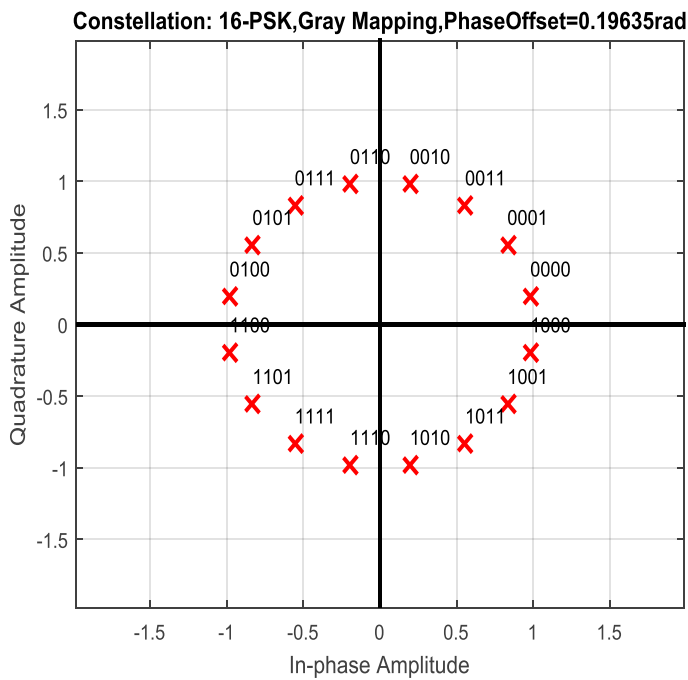

Fig.1. Constellation Diagram of 16-PSK

### B. M-ary QAM

QAM is a modulation technique of both analog and digital signal. It carries two different message signal (digital or analog) on the same carrier frequency, by changing amplitude and phase shifted by 90 degrees with respect to each other. The two independent signals are transmitted simultaneously over the same transmission medium. Tow signals are demodulated and the results are combined to produce the modulating signal at the receiver [8]. In the digital QAM case, a finite number of at least two phases and at least two amplitudes are used.

It can be expressed by equation (2)

$$S(t) = I_k \cos(\omega_c t) - Q_k \sin(\omega_c t) \ldots \ldots \ldots \ldots (2)$$

This equation shows the QAM signal in so-called IQ form that presents the modulation of the I-carrier and Q-carrier. It also can explain by,

$$S(t) = A_k \cos(\omega_c t + \phi_k) \quad \ldots \ldots \ldots \ldots (3)$$

Here,

$$A_k = \sqrt{I_k^2 + Q_k^2} \quad \ldots \ldots \ldots \ldots \ldots (4)$$
$$\phi_k = \tan^{-(Q_k/I_k)} \quad \ldots \ldots \ldots \ldots \ldots (5)$$

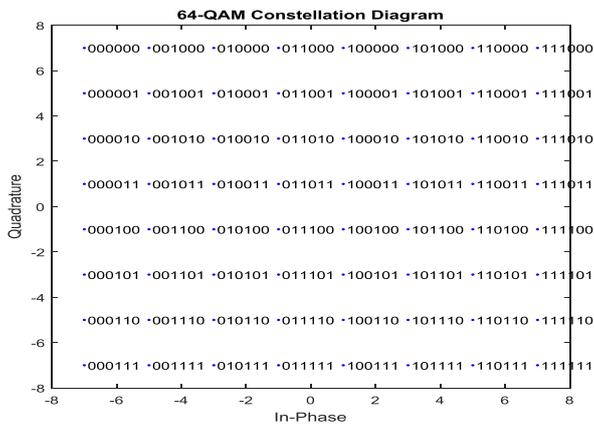

Fig.3. Constellation Diagram of 64-QAM

*C. Ricean fading channel*

It is a stochastic model for radio propagation variance affected by partial dissolution of a radio signal by itself. The signal reaches in the receiver by several different paths and at least one of the paths is altering or varying. It is useful for demonstrating mobile wireless communication systems when the transmitted signal can travel to the receiver along a leading line-of-sight or straight path. When one of the paths, usually a line of sight signal is much stronger than the others Rician fading occurs. It can be expressed by k which is the ratio between the power in the direct path and the power in the other paths.

This signal received over a Rician multipath channel can be expressed as

$$v(t) = C_{los} \cos \omega_c t + \Sigma^{N}_{i=1} \rho_i \cos(\omega_c t + \phi_i) \quad \ldots\ldots\ldots\ldots(6)$$

where,

$C_{los}$ = the amplitude of the line-of-sight element

$\rho_i$ = the amplitude of the n-th reflected wave

$\phi_i$ = the phase of the n-th reflected wave

i = 1,……, N identifies the reflected, scattered waves.

Rayleigh fading is recovered for $C_{los}= 0$.

*D. Bit Error Rate (BER)*

Bit error rate can be defined as the ratio of total number of error bit to the total number of transmitted bit. It is very essential way to govern the transmission superiority. It is often expressed as a percentage.

$$BER = \frac{Total\ number\ of\ error\ bits}{Total\ number\ of\ transmitted\ bits} \quad \ldots\ldots\ldots\ldots(7)$$

As an example, let the transmitted bit sequence:
0 1 0 1 0 0 0 1 1 0
And the following received bit sequence:
0 1 1 1 0 1 0 0 1 1
In this case the number of bit errors is 4 (Underlined symbol).

$BER = \frac{4}{10} \times 100\%$
$= 40\%$ error

*E. Signal to Noise Ratio (SNR)*

Signal to noise ratio means the ratio between the power of carrier signals to the power of noise signal in a wave. It is a measure used to find out the variation of the level of a desired signal with the level of corresponding noisy signal. It is expressed in logarithmic scale (dB). SNR also expressed by $\mu_0/2\mu$.

$$SNR = \frac{Signal\ Power}{Noise\ Power} \quad \ldots\ldots\ldots\ldots(8)$$

III. SIMULATION RESULTS

Figure 4 shows the BER performance of M-ary PSK for various values of $E_b/N_o$ (0 to 10 dB) in the case of Rician Fading Channel for fifth diversity order. Here the value of Bit error rate increases for a fixed $E_b/N_o$ with the increase of n ($M=2^n$) in M-ary PSK. For example in the simulation figure BER of 32 PSK is 0.2159 for SNR is 1dB where BER of 64 PSK is 0.2566 for same SNR. Table I and Table II explains that Large value of SNR causes less bit error rate and the value of BER of 64 PSK is comparatively larger than the 16 PSK. Figure 5 graphically represents the BER performance of M-ary QAM (M=4, 8,16,32,64,128,256,512,1024) for various values of SNR (0 to 10 dB) in the case of Rician Fading Channel when diversity order is 4.

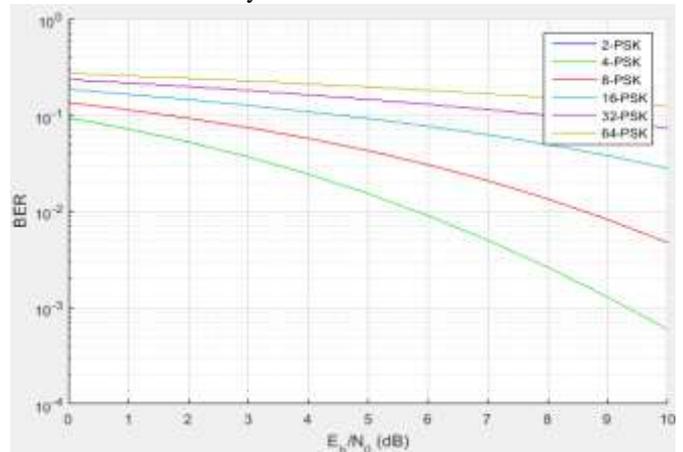

Fig.4. BER vs. $E_b/N_o$ curve of M-ary PSK for Rician Fading Channel when diversity order is 5.

Table I: BER OF 16-PSK FOR DIFFERENT EB/NO VALUES

| $E_b/N_o$(dB) | BER |
|---|---|
| 0 | 0.1856 |
| 1 | 0.1648 |
| 2 | 0.1449 |
| 3 | 0.1262 |
| 4 | 0.1086 |
| 5 | 0.0921 |
| 6 | 0.0768 |
| 7 | 0.0626 |
| 8 | 0.0496 |
| 9 | 0.0380 |
| 10 | 0.0280 |

TABLE II: BER OF 64-PSK FOR DIFFERENT EB/NO VALUES

| Eb/No(dB) | BER |
|---|---|
| 0 | 0.2718 |
| 1 | 0.2566 |
| 2 | 0.2414 |
| 3 | 0.2263 |
| 4 | 0.2112 |
| 5 | 0.1963 |
| 6 | 0.1814 |
| 7 | 0.1667 |
| 8 | 0.1521 |
| 9 | 0.1377 |
| 10 | 0.1237 |

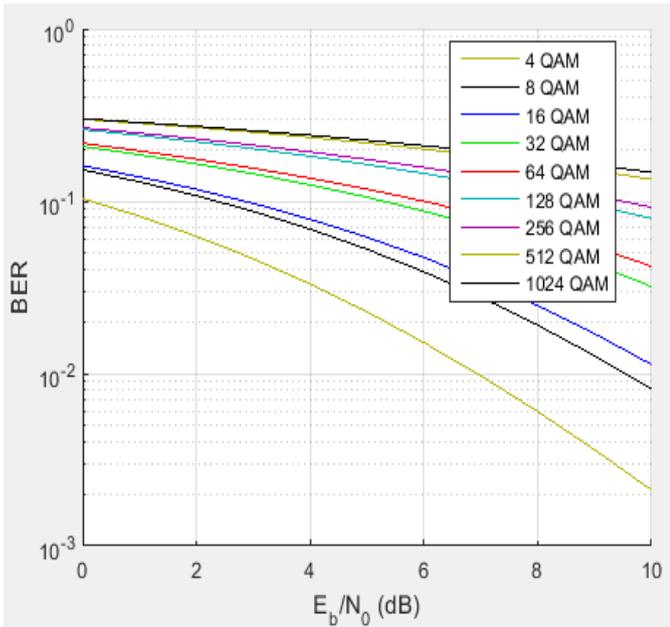

Fig.5. BER vs. SNR curve for M-ary QAM over Rician fading channel

TABLE III: BER OF 256-QAM FOR DIFFERENT EB/NO VALUES

| Eb/No(dB) | BER |
|---|---|
| 0 | 0.2616 |
| 1 | 0.2434 |
| 2 | 0.2248 |
| 3 | 0.2060 |
| 4 | 0.1872 |
| 5 | 0.1687 |
| 6 | 0.1507 |
| 7 | 0.1334 |
| 8 | 0.1169 |
| 9 | 0.1013 |
| 10 | 0.0865 |

TABLE IV: BER OF 512-QAM FOR DIFFERENT EB/NO VALUES

| Eb/No(dB) | BER |
|---|---|
| 0 | 0.2945 |
| 1 | 0.2789 |
| 2 | 0.2630 |
| 3 | 0.2466 |
| 4 | 0.2298 |
| 5 | 0.2128 |
| 6 | 0.1955 |
| 7 | 0.1782 |
| 8 | 0.1611 |
| 9 | 0.1444 |
| 10 | 0.1283 |

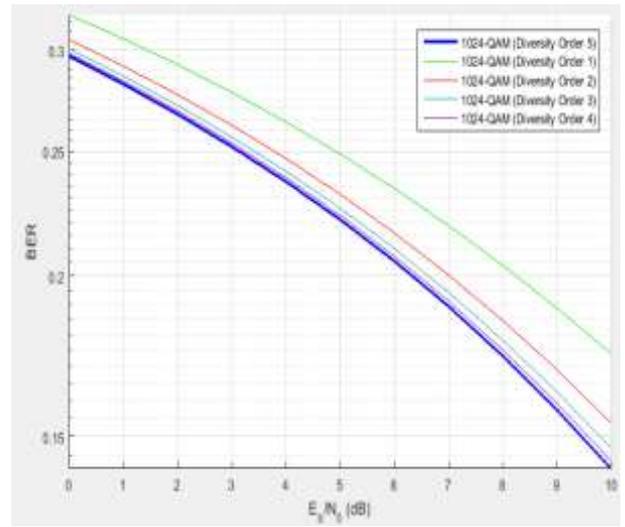

Fig.6. BER vs. SNR performance of M-ary QAM of Rician Fading Channel for different diversity order.

Table V: Diversity Order for Fixed Eb/No: 2db

| Diversity Order | BER (for 1024 QAM) |
|---|---|
| 1 | 0.2922 |
| 2 | 0.2768 |
| 3 | 0.2715 |
| 4 | 0.2689 |
| 5 | 0.2673 |

Table III and Table IV illustrates the BER performance of 256 QAM and 512 QAM respectively over the Rician Fading channel. Figure 6 graphically represents the BER vs. SNR performance of M-ary QAM for different diversity order. From Table V, it can be concluded that higher diversity order reduces the value of BER for 1024 QAM for a fixed SNR value of 2 dB.

## CONCLUSION

In this paper, two digital modulation schemes have been discussed theoretically and then implemented them in MATLAB simulation. QAM is a modulation technique of both analog and digital signal and M-ary PSK is the type of phase shift keying where variations of multiple phases are used instead of two phases. Discuss about Ricean fading channel with power factor and the performance of bit error rate in Ricean fading channel.